\documentclass[12pt]{article}

\usepackage{graphicx}
\usepackage{amsmath}
\usepackage{amssymb}

\usepackage{cite}

\def\CR{\nonumber \\}
\def\eq#1{(\ref{#1})}
\def\[#1\]{\begin{align}#1\end{align}}

\def\cH{{\cal H}}
\def\cJ{{\cal J}}
\def\tcH{{\tilde \cH}}
\def\tH{{\tilde H}}

\begin{document}

\begin{titlepage}
\title{
\hfill\parbox{4cm}{ \normalsize  YITP-15-52 \\ WITS-MITP-014}\\
\vspace{1cm} 
Constraint algebra of general relativity from a formal continuum limit of canonical tensor model}
\author{
Naoki {\sc Sasakura}$^a$\thanks{\tt sasakura@yukawa.kyoto-u.ac.jp} and Yuki {\sc Sato}$^b$\thanks{\tt Yuki.Sato@wits.ac.za}
\\[15pt]
$^a${\it Yukawa Institute for Theoretical Physics, Kyoto University,}\\
{\it Kyoto 606-8502, Japan} \\
\\
$^b${\it National Institute for Theoretical Physics, }\\
{\it School of Physics and Mandelstam Institute for Theoretical Physics,} \\
{\it University of the Witwatersrand, WITS 2050, South Africa}
}
\maketitle
\thispagestyle{empty}
\begin{abstract}
\normalsize
Canonical tensor model (CTM for short below) is a rank-three tensor model formulated as
a totally constrained system in the canonical formalism.
In the classical case, the constraints form a first-class constraint Poisson algebra 
with structures similar to that of the ADM formalism of general relativity,
qualifying CTM as a possible discrete formalism for quantum gravity. 
In this paper, we show that, 
in a formal continuum limit, the constraint Poisson algebra of CTM with no cosmological constant 
exactly reproduces that of the ADM formalism.
To this end, we obtain the expression of the metric tensor field in general relativity 
in terms of one of the dynamical rank-three tensors in CTM,
and determine the correspondence between the constraints of CTM and those of the ADM formalism.
On the other hand, the cosmological constant term of CTM seems 
to induce non-local dynamics, and is inconsistent with
an assumption about locality of the continuum limit.
\end{abstract}
\end{titlepage}

\section{Introduction}
\label{sec:introduction}
Tensor models \cite{Ambjorn:1990ge,Sasakura:1990fs,Godfrey:1990dt}
were originally introduced as models for quantum gravity in $D>2$ dimensions,
extending the matrix models which are considered to successfully describe quantum gravity in $D=2$. 
Subsequently, tensor models with group-valued indices \cite{Boulatov:1992vp,Ooguri:1992eb}, 
called group field theories \cite{DePietri:1999bx,Freidel:2005qe,Oriti:2011jm},
were introduced, which are especially studied in the context of loop quantum gravity.
The central idea of tensor models is that Feynman diagrams in tensor models 
may correspond to dual diagrams of discretized spacetimes.
Though the original models were not successful due to some difficulties 
\cite{Sasakura:1990fs, DePietri:2000ii},
colored tensor models \cite{Gurau:2009tw,Gurau:2011xp} with promising properties appeared, 
and have extensively been analyzed with interesting concrete 
results (see for instance \cite{Delepouve:2015nia,Bonzom:2015axa,Delepouve:2014hfa,Nguyen:2014mga,
Dartois:2014hga} for some recent developments). 
The colored tensor models also stimulated the renormalization group procedures of group field 
theories (see for instance \cite{Lahoche:2015ola,Benedetti:2014qsa,Geloun:2014ema,
Carrozza:2014rba} for recent developments).
There also appeared a new approach to random volumes in terms of matrix models
\cite{Fukuma:2015haa,Fukuma:2015xja}, which are in relation with colored tensor models.
  
The analysis of the colored tensor models has shown that leading orders of 
$1/N$\footnote{$N$ represents the number of discrete labels which an index takes: 
any index $a$ of tensors is assumed to take, say, $a=1,2,\ldots,N$, in this paper.}  expansions of the partition
functions are dominated by branched polymers
composed of melonic diagrams \cite{Gurau:2011xp,Bonzom:2011zz,Gurau:2013cbh}.
Naively, this would be an obstacle for a model of our spacetime, 
since branched polymers do not seem to represent extending entities 
like our real space, though there have been some interesting directions of study to change the situation by
considering higher orders \cite{Bonzom:2015axa,Raasakka:2013eda,Dartois:2013sra,
Kaminski:2013maa,Gurau:2013pca}. 
On the other hand,
it might be possible that the existence of a time-like direction is essentially important in quantum gravity, 
while the tensor models above basically deal with Euclidean signatures. 
This possibility arises from the fact that Causal Dynamical Triangulation 
has succeeded in generating de Sitter-like spacetimes \cite{Ambjorn:2004qm}, 
while the Euclidean cousin, Dynamical Triangulation, is not successful in 
this respect\footnote{When coupling many U$(1)$-fields, 
the authors in \cite{Horata:2000eg} found a promise of 
a phase transition higher than first order, 
which, however, is in conflict with the result in \cite{Ambjorn:1999ix}.}. 

Motivated by these considerations, one of the present authors has introduced 
a rank-three tensor model in Hamilton 
formalism \cite{Sasakura:2011sq,Sasakura:2012fb,Sasakura:2013gxg}
(There exists another Hamiltonian approach \cite{Oriti:2013aqa}
in the framework of group field theories.). 
Here the minimum choice of rank-three over matrices has been taken,
based on a belief that rank-three is enough to describe any dimensional space; 
this is a conclusion from the past works by one of 
the present authors on Euclidean rank-three tensor models
\cite{Sasakura:2008pe,Sasakura:2009hs}, 
though these models themselves are not successful due to serious necessities of fine-tuning. 
A time-like direction has been introduced by constructing Hamiltonian constraint forming a first-class 
constraint Poisson algebra with kinematical symmetries, which are the analog of   
spatial diffeomorphism in general relativity.
This way of introducing a time-like direction as gauge symmetry would be necessary for such a model
aiming for quantum gravity to reproduce general relativity in a (presently unknown) classical 
limit of continuous spacetime.
The requirement of the first-class nature of the constraint algebra
is so strong that the constraints and the algebraic structure of such a tensor model
are unique under some physically reasonable assumptions \cite{Sasakura:2012fb}.
Thus, our tensor model in the Hamilton formalism 
(canonical tensor model or CTM for short below) has turned out to be
formulated as a totally constrained system with first-class constraints, which have a Poisson algebraic 
structure very similar to the constraint Poisson algebra \cite{DeWitt:1967yk,Hojman:1976vp,
Teitelboim:1987zz} 
of the Arnowitt-Deser-Misner (ADM) formalism of general relativity \cite{Arnowitt:1960es,Arnowitt:1962hi}. 

The subsequent analyses have revealed some remarkable properties of CTM.
The $N=1$ case exactly reproduces the mini-superspace approximation of 
general relativity with a cosmological constant \cite{Sasakura:2014gia}. 
CTM can consistently be quantized \cite{Sasakura:2013wza},
and a number of exact physical wave functions for general $N$,
namely the exact solutions to the CTM analogue of the Wheeler-DeWitt equations, 
have been found \cite{Narain:2014cya}.
There is an intimate relation between CTM and statistical systems on random networks:
the Hamiltonian constraint of CTM generates the renormalization group flow
of randomly connected tensor networks \cite{Sasakura:2014zwa,Sasakura:2014yoa,Sasakura:2015xxa}.
This insight was remarkably useful in constructing the exact physical wave functions mentioned
above \cite{Narain:2014cya}.

The main purpose of the present paper is to consider a formal continuum limit of CTM
to find a relation with general relativity more general than the mini-superspace approximation
mentioned above.
We will show that, in the formal continuum limit, the first-class constraint Poisson algebra of CTM 
exactly agrees with that of the ADM formalism of general relativity
by properly taking into account a difference of weights between the Hamiltonian constraints
of the both theories.
The continuum limit contains the following two main assumptions:  
the indices of tensors can be replaced by continuous $D$-dimensional coordinates
with an implicit assumption of very large $N$,
and one of the dynamical tensors of CTM must have an almost diagonal form. 
Here, the off-diagonal components of the tensor are essentially important:
the lowest orders of a moment expansion for the off-diagonal components
will be identified with the (inverse) metric tensor field in general relativity.
It should be noted that we take the continuum limit in a formal manner,
and the limit must be justified by the dynamics of CTM in future study. 
We would also like to mention that a similar derivation of the constraint algebra of the ADM formalism
from that of CTM was done in a previous work \cite{Sasakura:2011sq}
by one of the present authors. The previous work, however, was obviously insufficient, 
because a specific Gaussian distribution of off-diagonal components
was assumed for computational simplicity, and coordinate dependences of variables 
were not fully accounted. On the contrary, the treatment of this paper is general and thorough.

This paper is organized as follows. 
In Section~\ref{sec:CTM}, the formalism of CTM is briefly recapitulated.
In Section~\ref{sec:CTMalgebra}, the first-class constraint Poisson algebra of CTM
is computed in the formal continuum limit.
Here, we introduce a moment expansion for the off-diagonal components of one of 
the dynamical tensors of CTM, and use the moments to express the continuum limit of the algebra. 
In Section~\ref{sec:interpretation}, we interpret the continuum limit
in terms of the ADM formalism of general relativity.  
This is successfully done by taking into account differences of weights 
of the gauge parameters between CTM and the ADM formalism, and 
by introducing an assumed relation between the lowest orders of the moment expansion
and the (inverse) metric tensor field.
In Section~\ref{sec:modified}, we consider the Hamiltonian constraint 
obtained by multiplying that of the ADM formalism by a weight of half-density, 
and study the constraint Poisson algebra among the newly defined Hamiltonian constraint
and the momentum constraint.
The algebra certainly reproduces the continuum limit of that of CTM, and hence this 
proves the equivalence of the two.
The final section is devoted to summary and discussions. 

\section{Canonical tensor model}
\label{sec:CTM}
The set of the dynamical variables of the canonical tensor model (CTM)  
\cite{Sasakura:2011sq,Sasakura:2012fb,Sasakura:2013gxg}
in the minimal setting \cite{Sasakura:2013wza}
is given by a canonical conjugate pair of symmetric real rank-three tensors, 
$M_{abc}$ and $P_{abc}$ ($a,b,c=1,2,\ldots,N$), satisfying
\[
\{ M_{abc}, P_{def} \} 
=\sum_{\sigma}
 \delta_{a\sigma_d} \delta_{b \sigma_e} \delta_{c \sigma_f},
 \ \ 
 \{ M_{abc}, M_{def} \}=\{P_{abc}, P_{def} \}=0, 
 \label{eq:poisson}
\] 
where $\{\ ,\ \}$ denotes the Poisson bracket, and the summation is over all the permutations 
of $d,e,f$ to incorporate the symmetry of the tensors under all the permutations of the indices.
The Hamiltonian is given by
\[
H=\xi_a \cH_a +\eta_{[ab]} \cJ_{[ab]}, 
\label{eq:hamiltonian}
\] 
where $\xi_a$ and $\eta_{[ab]}$ are Lagrange multipliers, repeated indices are summed over, and 
\[
&\cH_a = \frac{1}{2} \left( P_{abc} P_{bde} M_{cde} 
- \lambda M_{abb}  \right), \label{eq:defofH} \\
&\cJ_{[ab]} = \frac{1}{4} \left(P_{acd}M_{bcd} 
- P_{bcd} M_{acd}  \right) \label{eq:defofJ}.
\]
Here the square brackets in the indices symbolically represent the anti-symmetry, $\cJ_{[ab]}=-\cJ_{[ba]},
\ \eta_{[ab]}=-\eta_{[ba]}$,
and $\cJ_{[ab]}$ and $\cH_a$ are the generators of the $SO(N)$-kinematical symmetry
and of the symmetry analogous to the temporal 
diffeomorphism in general relativity, respectively.  
Following the naming in the ADM formalism \cite{Arnowitt:1960es,Arnowitt:1962hi} 
of general relativity, 
we call $\cH_a$ and $\cJ_{[ab]}$ \textit{Hamiltonian constraint}  and \textit{momentum constraint}
of CTM, respectively. 
$\lambda$ is a real undetermined constant, which we call \textit{cosmological constant}.  
The last naming comes from the fact that 
the $N=1$ case exactly agrees with the mini-superspace approximation of 
general relativity with a cosmological constant proportional to $\lambda$ \cite{Sasakura:2014gia}.

As in the case of the ADM formalism, $\cH_a$ and $\cJ_{[ab]}$ form a first-class constraint Poisson algebra
 given by 
\[
&\{\cH (\xi^1), \cH (\xi^2)\} =\cJ ( [\tilde \xi^1, \tilde{\xi^2} ] 
+ 2 \lambda [[\xi^1, \xi^2]] ), \notag \\
&\{\cJ (\eta), \cH (\xi)\}= \cH \left(\eta\, \xi \right), 
\label{eq:constraintalg}\\
&\{\cJ (\eta^1), \cJ (\eta^2)\} =  \cJ \left( [\eta^1,\eta^2] \right), \notag
\]
where $\cH(\xi)\equiv \xi_a\cH_a$, $\cJ (\eta)\equiv \eta_{[ab]}\cJ_{[ab]}$, 
$\tilde{\xi}_{ab}\equiv P_{abc}\xi_c$. 
Here, on the right-hand sides, $[\ ,\ ]$ denotes the matrix commutator,
and $[[\xi^1,\xi^2]]_{ab}\equiv \xi^1_a \xi^2_b-\xi^2_a \xi^1_b$. 
It is important to note that the algebra \eq{eq:constraintalg}
has a structure depending on $P$ on the right-hand side in the first line,
and therefore it is not a genuine Lie algebra. 
This is a similar situation as 
the constraint algebra \cite{DeWitt:1967yk,Hojman:1976vp,Teitelboim:1987zz} of general relativity 
in the ADM formalism, and will be essential for \eq{eq:constraintalg} to reproduce 
the constraint algebra of the ADM formalism in a formal continuum limit.
The constraints \eq{eq:defofH}, \eq{eq:defofJ} and hence the first-class algebra \eq{eq:constraintalg} 
are unique under some physically reasonable assumptions \cite{Sasakura:2012fb}.

\section{Constraint algebra of CTM in a formal continuum limit}
\label{sec:CTMalgebra}
In this section, we take a continuum limit of the constraint algebra \eq{eq:constraintalg} of CTM 
by assuming emergence of a continuum space. 
Note that this assumption is imposed without any justifications. 
Ideally, such an assumption should be derived as infrared effective dynamics of CTM, 
but this is out of our reach at present. Namely, the derivation in this section should be 
considered as a formal continuum limit ignoring the presently unknown real dynamics of CTM. 
On the other hand, it will suggest a plausible way for emergence of space and general relativity
in the framework of CTM, and give directions of future study to finally justify the assumption.  

First of all, the assumption will be translated to that the indices of the 
tensors can be replaced by continuous coordinates of {\it space} as
\[
a \rightarrow x\in {\rm R}^D,
\]
where $D$ denotes the spatial dimension.
Then, we also assume that index contractions are replaced by integrations as
\[
\sum_{a=1}^N \rightarrow \int d^Dx.
\] 
In general, there could exist a 
non-trivial integration measure as $\int d^Dx\, \rho(x)$, but this could be canceled by 
a Jacobian $|\frac{\partial x'}{\partial x}|$ after an appropriate transformation of the coordinate $x'(x)$.
Note that there are no contradictions in regarding $x$ to be fixed labels as here, 
since the transformations below apply to the dynamical variables of CTM but not to the indices represented
by the coordinates.

A continuum space has an intrinsic concept of locality, and we will pick up a local part of 
the constraint algebra \eq{eq:constraintalg}. 
We will also introduce an assumption about the form of $P$ which is in accord with the locality.
We will show that this reduction to a local part of the algebra can consistently be done,
except for the cosmological constant term in \eq{eq:defofH}.
We will comment on the cosmological constant term in the final section.

\subsection{$\{\cJ,\cJ\}$ part}
Let us first discuss the Poisson algebra of $\cJ$. 
The local part of the algebra with respect to the continuum space 
would be picked up by putting the $D$-dimensional delta-function 
$\delta^D(x-y)$ into the argument of $\cJ(\eta)$ in some manner. 
Since $\eta$ must be anti-symmetric,
the lowest order should be expressed by using the first derivatives of $\delta^D(x-y)$, and  
we are lead to the form, 
\[
\eta_{[xy]}=\frac{1}{2} \left( v^\mu(x)+v^\mu(y) \right) \delta^D_\mu (x-y),
\label{eq:etadelta}
\]
where $v$ is an arbitrary vector field on the space, the Greek index $\mu$ denotes the spatial 
indices, $\mu=1,2,\ldots,D$, and 
\[
\delta^D_\mu (x-y)\equiv \frac{\partial}{\partial x^\mu} \delta^D(x-y). 
\label{eq:deltamu}
\]

The third equation of \eq{eq:constraintalg} implies that the algebra of $\cJ$ is equivalent to 
the commutator algebra of $\eta$. 
Products of distributions can be computed rather easily by using test functions. 
In the present case, for a test function $f$, we obtain
\[
(\eta f) (x)&\equiv 
\int d^Dy\, \eta_{[xy]} f(y) \CR
&=\frac{1}{2} \int d^Dy \, \left( v^\mu(x)+v^\mu(y) \right) \delta^D_\mu (x-y) f(y) \CR
&= \frac{1}{2} \int d^Dy \, \delta^D (x-y)\frac{\partial}{\partial y^\mu}
\left( \left( v^\mu(x)+v^\mu(y) \right) f(y)\right) \CR
&= \left( \frac{1}{2} (\partial_\mu v^\mu)  +v^\mu \partial_\mu \right)f(x). 
\label{eq:operatef}
\]
So, we obtain
\[
[\eta^1,\eta^2]f&=\eta^1 (\eta^2 f)-\eta^2 (\eta^1 f)\CR 
&=\left( \frac{1}{2} (\partial_\mu v^\mu_1)+ v_1^\mu \partial_\mu \right)
\left(\frac{1}{2} (\partial_\mu v^\mu_2 )f +v_2^\mu \partial_\mu \right)f
 - (1\leftrightarrow 2 )  \CR
&=\left( \frac{1}{2} (\partial_\mu v^\mu_3)  +v_3^\mu \partial_\mu \right)f,
\label{eq:cometa}
\]
where $v_{1,2}$ are respectively related to $\eta^{1,2}$ as \eq{eq:etadelta}, and 
\[
v_3^\mu=[v_1,v_2]^\mu=v_1^\nu \partial_\nu v_2^\mu-v_2^\nu \partial_\nu v_1^\mu.
\label{eq:v1v2}
\]
Note that \eq{eq:cometa} has exactly the same form as \eq{eq:operatef}.
Therefore, 
\[
[\eta^1,\eta^2]=\eta^3,
\]
where $\eta^3$ is \eq{eq:etadelta} with $v=v_3$.
Thus, the commutator algebra of $\eta$ with the form \eq{eq:etadelta} closes, and 
the representative vectors $v$ form a commutation algebra given in \eq{eq:v1v2}.

\subsection{ $\{ \cJ,\cH \}$ part}
Under the assumption of the continuum limit, 
the vector $\xi$ of $\cH(\xi)$ should be replaced by a function $\xi(x)$ on the $D$-dimensional space. 
Then, the second line of \eq{eq:constraintalg} implies that the Poisson bracket between $\cJ$ and $\cH$ 
is equivalent to the matrix-vector product $\eta\, \xi$. 
Thus, by using \eq{eq:etadelta} and \eq{eq:operatef}, we obtain
\[
\eta\, \xi=\frac{1}{2} (\partial_\mu v^\mu) \xi  +v^\mu \partial_\mu \xi,
\label{eq:etaxi}
\]
where $v$ is the representative vector of $\eta$.

\subsection{ $\left \{ \cH, \cH  \right\}$ part} 
This is the most non-trivial part of our discussions, and
we have to take into account not only the leading order in the expansion regarding locality
but also the next leading orders 
to obtain a non-vanishing result.
The next leading orders are dimensionally higher than the leading one by (length)$^2$.
This would mean that $\left \{ \cH, \cH  \right\}$ part contains an effective length scale, and 
the way it appears suggests that it is intimately related with the smallest length scale 
in a space or the Planck length.
In this paper, its definitive interpretation is beyond our scope, and is left for future study.  

Let us first consider the part independent of the cosmological constant $\lambda$,
namely $[\tilde \xi^1,\tilde \xi^2]$ on the right-hand side in the first line of \eq{eq:constraintalg}. 
Let us first try a strictly local form of $P$, $P^\delta_{xyz}\equiv \delta^D(x-y)\delta^D(x-z)$,
and evaluate  $[\tilde \xi^1,\tilde \xi^2]$.
The operation $\tilde \xi$ on a test function $f$ can be computed as 
\[
(\tilde \xi f)(x)&=\int d^Dyd^Dz P^\delta_{xyz}\,\xi(y) f(z) \CR
&= (\xi f)(x),
\]
and hence
\[
[\tilde \xi^1,\tilde \xi^2]f&=\tilde \xi^1(\tilde \xi^2 f)-\tilde \xi^2(\tilde \xi^1 f) \CR
&=\xi^1 \xi^2 f-\xi^2 \xi^1 f \CR
&=0.
\]
Therefore, $P$ needs to be smeared for $[\tilde \xi^1,\tilde \xi^2]$ to be non-vanishing.

To characterize such smearing of $P$ in the order of 
(length)$^2$\footnote{As shown below, $\beta$ looks like in the first order in \eq{eq:moments},
but turns out to be in the second order due to the permutation symmetry of $P$.}, 
let us introduce the following moments,
\[
&\int d^Dy d^Dz P_{xyz}=\alpha(x), \CR
&\int d^Dy d^Dz P_{xyz} \delta y^\mu=\int d^Dy d^Dz P_{xyz} \delta z^\mu=\beta^\mu(x),\CR
&\int d^Dy d^Dz P_{xyz}\delta y^\mu \delta y^\nu=\int d^Dy d^Dz P_{xyz}\delta z^\mu \delta z^\nu
=\gamma^{\mu\nu}(x), 
\label{eq:moments} \\
&\int d^Dy d^Dz P_{xyz}\delta y^\mu \delta z^\nu=\tilde \gamma^{\mu\nu}(x), \nonumber
\]
where $\delta y=y-x,\ \delta z=z-x$, and we have taken into account 
$P_{xyz}=P_{xzy}$, which is one of the permutation symmetries of $P$. 
In fact, the remaining permutation symmetry of $P$ gives more restrictions on 
the moments in \eq{eq:moments}. 
To see this, let us consider three independent test functions $f_i\ (i=1,2,3)$.
By using \eq{eq:moments} and
performing partial integrations, we obtain
\[
&\int d^Dx d^Dy d^Dz\, P_{xyz} f_1(x)f_2(y)f_3(z)\CR
&=\int d^Dxd^Dyd^Dz\, P_{xyz} f_1
\left( f_2+\delta y^\mu f_{2,\mu}+\frac{1}{2} \delta y^\mu \delta y^\nu f_{2,\mu\nu} \right) 
\left( f_3+\delta z^\mu f_{3,\mu}+\frac{1}{2} \delta z^\mu \delta z^\nu f_{3,\mu\nu} \right)+\cdots \CR
&=\int d^Dx \left( \alpha f_1f_2f_3+ \beta^\mu f_1(f_2f_3),_{\mu} 
+ \tilde \gamma^{\mu\nu} f_1f_{2,\mu} f_{3,\nu}+\frac{1}{2}\gamma^{\mu\nu} f_1 f_2 f_{3,\mu\nu} 
+\frac{1}{2} \gamma^{\mu\nu} f_1 f_{2,\mu\nu}f_3\right)+\cdots
 \CR
&=\int d^Dx \Bigg( 
\left( \alpha-\beta^\mu_{,\mu}+\frac{1}{2} \gamma^{\mu\nu}_{,\mu\nu}\right) f_1f_2f_3
+\left( \gamma^{\mu\nu}_{,\nu} -\beta^\mu \right) f_{1,\mu}f_2f_3
+\left(\gamma^{\mu\nu}_{,\nu}-\tilde \gamma^{\mu\nu}_{,\nu}\right) f_1f_2f_{3,\mu}\CR
&\hspace{2cm}-f_2 \tilde  \gamma^{\mu\nu} \left( f_1 f_{3,\mu\nu}+f_{1,\mu}f_{3,\nu}\right) 
+\frac{1}{2}f_2  \gamma^{\mu\nu}\left( 2 f_1f_{3,\mu\nu}+f_{1,\mu\nu}f_3+2 f_{1,\mu}f_{3,\nu}\right)
\Bigg)+\cdots,
\label{eq:permfint}
\]
where we have expanded the test functions $f_2(y),f_3(z)$ around $x$, 
suppressing the obvious argument $x$,
and have used the shorthand notations, $f_{,\mu}=\partial_\mu f$, etc.
Here the dots represent terms with higher moments and will be ignored.
The third line contains no derivatives of $f_1$, and, from the third line to the final ones, we have 
performed partial integrations to remove the derivatives of $f_2$.
Since the permutation symmetry of $P$ requires that \eq{eq:permfint} must 
be symmetric under the permutations of the test functions, the third and the final lines 
should have the same expression after the exchange of $f_1$ and $f_2$.
This demands
\[
&\alpha-\beta^\mu_{,\mu}+\frac{1}{2} \gamma^{\mu\nu}_{,\mu\nu}=\alpha, \CR
&\gamma^{\mu\nu}_{,\nu} -\beta^\mu
=\gamma^{\mu\nu}_{,\nu}-\tilde \gamma^{\mu\nu}_{,\nu}=\beta^\mu, \CR
&\gamma^{\mu\nu}-\tilde \gamma^{\mu\nu}=\tilde \gamma^{\mu\nu}, \\
&\gamma^{\mu\nu}-\tilde \gamma^{\mu\nu}=\frac{1}{2}\gamma^{\mu\nu}, \nonumber
\]
and the solution is
\[
&\beta^\mu=\frac{1}{2} \gamma^{\mu\nu}_{,\nu}, \CR
&\tilde \gamma^{\mu\nu}=\frac{1}{2}\gamma^{\mu\nu}.  
\label{eq:restrictions}
\]
Thus we obtain that the moment expansion \eq{eq:moments} is characterized thoroughly 
by $\alpha$ and $\gamma$.
Here, it is important to recall that we are {\it not} performing derivative expansions
for dynamical variables, but rather we are performing expansions in moments of ``fuzziness" of space:
we count the number of derivatives on test functions as orders, but not derivatives
on dynamical variables such as $\alpha,\gamma$.
Therefore, the orders associated to $\alpha$ and $\gamma$ are zeroth and second, respectively, 
and $\beta$ should be considered to be in the same order as $\gamma$,
irrespective of a derivative on $\gamma$ in \eq{eq:restrictions}.
 
By using \eq{eq:moments} and \eq{eq:restrictions}, 
the operation $\tilde \xi$ in \eq{eq:constraintalg} on a test function $f$ can be computed as 
\[
(\tilde \xi f)(x)&=\int d^Dy d^Dz P_{xyz} \xi(y) f(z)  \CR
&= \int d^Dy d^Dz P_{xyz}\left( \xi(x)+\delta y^\mu \xi_{,\mu}(x)+\frac{1}{2} \delta y^\mu \delta y^\nu 
\xi_{,\mu\nu}(x)\right) 
\CR
&\hspace{3cm} \times 
\left( f(x)+\delta z^\mu  f_{,\mu}(x)+\frac{1}{2} \delta z^\mu \delta z^\nu 
f_{,\mu\nu}(x)\right)+\cdots \CR
&=\left( \alpha \xi f +\frac{1}{2} \gamma^{\mu\nu}_{,\mu} (\xi f)_{,\nu}+
\frac{1}{2} \gamma ^{\mu\nu}\left(\xi_{,\mu} f_{,\nu} + \xi f_{,\mu\nu}+f \xi_{,\mu\nu}\right)  \right)(x)+\cdots.
\label{eq:xif}
\]
Then, by using \eq{eq:xif}, we obtain the commutator $[\tilde \xi^1,\tilde \xi^2]$ as
\[
[\tilde \xi^1,\tilde \xi^2]f&=\tilde \xi^1(\tilde \xi^2 f)-\tilde \xi^2(\tilde \xi^1 f) \CR
&= \frac{1}{2} \alpha \xi^1 \gamma^{\mu\nu}_{,\mu} \xi^2_{,\nu}f+
\frac{1}{2}\gamma^{\mu\nu} \left( \alpha\xi^1 \xi^2_{,\mu\nu} f+2 \alpha \xi^1 \xi^2_{,\mu} f_{,\nu}
+ \xi^1 \alpha_{,\mu} \xi^2_{,\nu} f  \right) - (1\leftrightarrow 2),
\label{eq:comxixi}
\]
where we have discarded terms with orders $(\gamma)^2$ and higher. 
In \eq{eq:constraintalg}, $[\tilde \xi^1,\tilde \xi^2]$ is the argument of the generator $\cJ$,
and, surprisingly, \eq{eq:comxixi} has exactly the form of \eq{eq:operatef} with
\[
v^\mu=\alpha \gamma^{\mu\nu} \left( \xi^1 \xi^2_{,\nu}- \xi^2 \xi^1_{,\nu}\right).
\label{eq:vxirel}
\]
This shows that the local form of $\eta$ in \eq{eq:etadelta} is consistent with 
the algebra of the Hamiltonian constraints 
in the second order of the moment expansion of $P$,
except for the part depending on the cosmological constant $\lambda$.

As for the term depending on the cosmological constant $\lambda$ 
on the right-hand side in the first equation of \eq{eq:constraintalg}, 
we obtain
\[
\left[\left[ \xi^1,\xi^2 \right]\right]_{xy}=\xi^1(x)\xi^2(y)- \xi^2(x)\xi^1(y).
\label{eq:cos}
\]
Since we may take any $x,y$, the expression is generally non-local
with respect to the continuum space.

\section{Interpretation in terms of geometrodynamics}
\label{sec:interpretation}
In Section~\ref{sec:CTMalgebra}, we have analyzed the structure of 
the constraint algebra of CTM on the assumption of the emergence of a continuous 
$D$-dimensional space with its intrinsic locality.
In this section, we will compare the result with the constraint 
algebra \cite{DeWitt:1967yk,Hojman:1976vp,Teitelboim:1987zz}
of the ADM formalism \cite{Arnowitt:1960es,Arnowitt:1962hi} of 
general relativity, and will construct an exact correspondence 
except for the part proportional to the cosmological constant $\lambda$.

In the Hamilton formalism, general relativity can be formulated as a totally constrained system 
due to the gauge symmetry of the spacetime diffeomorphism.
In the ADM formalism, the Hamiltonian is given by 
\[
H^{ADM}=\int d^Dx \left( n(x) \cH^{GR}(x)+ w^\mu(x) \cJ^{GR}_\mu(x) \right),
\]
where $\cH^{GR}$ and $\cJ_\mu^{GR}$ are respectively 
the Hamiltonian and momentum constraints, 
and $n$ and $w^\mu$ are the lapse function and the shift vector, respectively.
The constraints satisfy the first-class constraint Poisson algebra,
\[
&\{ H^{GR}(n_1),H^{GR}(n_2) \}=J^{GR}(\tilde n), \CR
&\{ J^{GR}(w),H^{GR}(n) \}=H^{GR} ({\cal L}_w n), \label{eq:grpoisson}\\
&\{ J^{GR}(w_1),J^{GR} (w_2) \} =J^{GR} ({\cal L}_{w_1} w_2), \nonumber
\]
where
\[
&H^{GR}(n)=\int d^Dx\ n(x)\, \cH^{GR}(x), 
\label{eq:defofHGR}\\
&J^{GR}(w)=\int d^Dx\ w^\mu(x)\, \cJ^{GR}_\mu(x),
\label{eq:defofJGR} \\
&\tilde n^\mu=g^{\mu\nu} (n_1 \partial_\nu n_2-n_2 \partial_\nu n_1), 
\label{eq:ntilde}
\]
and ${\cal L}$ denotes the Lie derivative,
\[
&{\cal L}_w n= w^\mu\partial_\mu n, \label{eq:wn}\\
&({\cal L}_{w_1} w_2)^\mu=[w_1,w_2]^\mu=w_1^\nu\partial_\nu w_2^\mu-
w_2^\nu\partial_\nu w_1^\mu.
\label{eq:w1w2}
\]

An important feature of \eq{eq:grpoisson} is that the algebraic structure depends on the 
dynamical field, the spatial inverse metric $g^{\mu\nu}(x)$, as in \eq{eq:ntilde},
and the algebra is therefore not a genuine Lie algebra. 
This structure is an essence of  geometrodynamics, as thoroughly discussed  in \cite{Hojman:1976vp}.
Dependence on the dynamical variable $P$ exists similarly in the constraint algebra 
\eq{eq:constraintalg} of CTM, and was important in deriving the constraint algebra in 
the continuum limit of CTM in Section~\ref{sec:CTMalgebra}. 

Let us first compare $\{{\cal J},{\cal J}\}$ with $\{J^{GR},J^{GR}\}$. By comparing \eq{eq:v1v2} with
\eq{eq:w1w2}, we can simply identify
\[
v=w.
\label{eq:v=w}
\]
Thus $\cJ(\eta)$ with \eq{eq:etadelta} of CTM can be identified with the spatial diffeomorphism 
in the continuum limit.

Next, let us compare $\{ \cJ,\cH\}$ with $\{ J^{GR}, H^{GR}\}$. 
With the identification \eq{eq:v=w},
it is not possible to identify $\xi$ and $n$, because there is a difference of weights between \eq{eq:etaxi} 
and \eq{eq:wn}. Geometrically, $n$ is a scalar, but $\xi$ is a scalar half-density. 
The difference can be balanced by assuming
\[
\xi=g^\frac{1}{4} n,
\label{eq:xi=n}
\]
where $g=\hbox{Det}[ g_{\mu\nu}]$. In fact, by assuming \eq{eq:wn} and using
${\cal L}_w g_{\mu\nu}=\nabla_\mu w_\nu+\nabla_\nu w_\mu$, one obtains
\[
{\cal L}_w (g^A n)&=2 A (\nabla_\mu w^\mu) g^A n+
g^A w^\mu \partial_\mu n \CR
&=2 A (\partial_\mu w^\mu+\Gamma^{\nu}_{\nu\mu}w^\mu ) g^A n+
g^A w^\mu \partial_\mu n  \CR
&=2 A (\partial_\mu w^\mu) g^A n+
w^\mu \partial_\mu (g^A n),
\label{eq:lietrans}
\]
where $A$ is a number.
For $A=\frac{1}{4}$, \eq{eq:lietrans} agrees with \eq{eq:etaxi} under \eq{eq:v=w}, 
and therefore we should choose as \eq{eq:xi=n}.

Lastly, we compare $\{\cH,\cH\}$ with $\{H^{GR},H^{GR}\}$. By putting \eq{eq:xi=n}
into \eq{eq:vxirel}, we obtain
\[
v^\mu&=\alpha \gamma^{\mu\nu} g^\frac{1}{4} \left( n_1 \partial_\nu (g^\frac{1}{4} n_2) 
-n_2 \partial_\nu (g^\frac{1}{4} n_1) \right) \CR
&= \alpha \gamma^{\mu\nu} g^\frac{1}{2} \left( n_1 \partial_\nu n_2-n_2 \partial_\nu n_1 \right).
\]
By comparing this with \eq{eq:ntilde} under \eq{eq:v=w}, we obtain
\[
\alpha \gamma^{\mu\nu} =g^{-\frac12}g^{\mu\nu}.
\label{eq:gamma=g}
\] 

As for the the part proportional to the cosmological constant on the righthand side 
in the first equation of \eq{eq:constraintalg}, 
it seems difficult to associate a geometrodynamical interpretation
due to the non-local character shown in \eq{eq:cos}. 

Finally, we will make a comment on the weights appearing in \eq{eq:xi=n} and \eq{eq:gamma=g}. 
As shown in \eq{eq:operatef}, with the identification \eq{eq:v=w},
a vector in CTM is translated to a quantity which transforms as a scalar half-density by
$\cJ(\eta)$ with \eq{eq:etadelta}. This applies to $\xi(x)$ as in \eq{eq:etaxi}
as well as to each index of $P$. Therefore, we make assignments,
\[
&[\xi]_w=\frac{1}{2}, 
\label{eq:weightxi} \\
&[P]_{w}=\frac{3}{2},
\label{eq:weightP}  
\]
where we have introduced $[\ ]_w$ to denote the weight of a quantity; 
the weight is defined so that a quantity $q$ transforms as a scalar density, if $[q]_w=1$. 
On the other hand, we have
\[
\left[ \int d^Dx \right]_w=-1, 
\label{eq:weightint}
\] 
since an integration over the space cancels the transformation of a quantity $q(x)$ with 
$[q]_w=1$ as
\[
\int d^D x \left( \left(\partial_\mu w^\mu\right) q+ w^\mu \partial_\mu q\right)
=\int d^Dx\, \partial_\mu (w^\mu q)=0, 
\] 
where we have ignored possibilities of boundary contributions.
By applying \eq{eq:weightP} and \eq{eq:weightint} to the quantities in \eq{eq:moments}, we obtain
\[
\left[ \alpha \right]_w=\left[ \gamma^{\mu\nu} \right]_w=-\frac{1}{2},
\]
and therefore $\left[ \alpha \gamma^{\mu\nu} \right]_w=-1$.
This explains the power of $g$ in \eq{eq:gamma=g}.

\section{Modified constraint algebra in the ADM formalism}
\label{sec:modified}
In Section~\ref{sec:interpretation}, it has been shown that, to relate CTM to the ADM formalism,  
the gauge parameters $\xi$ and $n$ associated respectively to 
the Hamiltonian constraints of CTM and the ADM formalism should have different weights as in \eq{eq:xi=n}. 
In the discussions, a factor $g^\frac{1}{4}$ 
has been introduced to modify the wight of the gauge parameter $n$, 
and $g_{\mu\nu}(x)$ in it was treated as a metric tensor field on the space. 
In the Hamilton formalism, however, $g_{\mu\nu}(x)$ is a dynamical variable, 
and therefore the wight factor should be considered to be a modification of the Hamiltonian 
constraint rather than the gauge parameter.
Then, the additional weight factor multiplied on the Hamiltonian constraint 
may potentially ruin the correspondence argued in
Section~\ref{sec:interpretation} on account of the Poisson brackets between the 
weight factor and the constraints.
Therefore, in this section, we will explicitly write down the constraint algebra after the change of   
the weight of the Hamiltonian constraint in the ADM formalism, 
and will show that it actually agrees with the continuum limit of 
the constraint algebra of CTM, namely \eq{eq:etaxi} and \eq{eq:vxirel} with \eq{eq:gamma=g} 
(Since there is no change in the momentum constraint, \eq{eq:v1v2} does not need to be checked.).    
 
Let us define the modified Hamiltonian constraint $\tcH$ (and $\tH$) as 
\[
&\tcH=g^{B} \cH^{GR}, \\
&\tH(\xi)=\int d^Dx  \, \xi(x) \tcH (x), 
\]
where $g={\rm Det}(g_{\mu\nu})$ and $B$ is a number. Here, we have changed the weight of the Hamiltonian constraint,
and, from \eq{eq:gamma=g}, we expect that the constraint algebra of CTM can be obtained
for $B=-\frac{1}{4}$.  

The Hamiltonian constraint of the ADM formalism satisfies
\[
\{ g_{\mu\nu}(x), \cH^{GR}(y) \}= C_{\mu\nu}(x)\delta^D(x-y),
\label{eq:gch}
\]
where $C_{\mu\nu}(x)=-2 K_{\mu\nu}(x)$, being proportional to the extrinsic curvature 
expressed in terms of $g_{\mu\nu}(x)$ and its conjugate momentum $\pi^{\mu\nu}(x)$
\cite{DeWitt:1967yk,Hojman:1976vp,Teitelboim:1987zz}. 
Here, what matters in the following discussion is only the fact that the right-hand side of \eq{eq:gch}
is strictly local: it does not contain derivatives of $\delta^D(x-y)$. 
From \eq{eq:gch}, we obtain
\[
\{ g(x)^B,\cH^{GR}(y) \}= B g(x)^B C(x) \, \delta^D(x-y),
\label{eq:strict}
\] 
where $C=g^{\mu\nu} C_{\mu\nu}$.
Then, we find that
\[
\{ \tcH (x),\tcH(y)  \}&=\{ g(x)^B \cH^{GR}(x), g(y)^B \cH^{GR}(y)\} \CR
&=g(x)^B\cH^{GR}(y)\{  \cH^{GR}(x), g(y)^B \} 
+g(y)^B \cH^{GR}(x)\{ g(x)^B , \cH^{GR}(y)\} \CR
&\ \ \ \ +g(x)^Bg(y)^B \{  \cH^{GR}(x),  \cH^{GR}(y)\} \CR
&=g(x)^Bg(y)^B \{  \cH^{GR}(x),  \cH^{GR}(y)\},
\label{eq:tcHtcH}
\]
where the two terms in the second line have canceled with each other
due to the strict local form of \eq{eq:strict}.
Therefore, by using \eq{eq:grpoisson}, \eq{eq:defofHGR}, 
\eq{eq:defofJGR}, \eq{eq:ntilde}, and \eq{eq:tcHtcH}, we obtain
\[
\{ \tH(\xi^1),\tH(\xi^2)\}&=\int d^Dxd^Dy\ \xi^1(x) \xi^2(y) \{ \tcH (x),\tcH(y)  \} \CR
&=\int d^Dx d^Dy\ g(x)^B\xi^1(x)g(y)^B \xi^2(y) \{  \cH^{GR}(x),  \cH^{GR}(y)\} \CR
&=\int d^Dx \ g^{\mu\nu}(x) \left( g(x)^B \xi^1(x) \partial_\nu (g(x)^B \xi^2(x)) -
(1 \leftrightarrow 2) \right) \cJ^{GR}_\mu(x) \CR
&=J^{GR}(v),
\] 
where
\[
v^\mu=g^{\mu\nu} g^{2B} (\xi^1 \partial_\nu \xi^2- \xi^2 \partial_\nu \xi^1).
\]
Indeed, for $B=-\frac{1}{4}$, this agrees with \eq{eq:vxirel} under \eq{eq:gamma=g}. 

We will next compute $\{ J^{GR}(v), \tH(\xi)\}$. Let us first remind the explicit expression of
the momentum constraint \cite{DeWitt:1967yk,Hojman:1976vp,Teitelboim:1987zz}
in the ADM formalism,
\[
\cJ^{GR}_\mu=-2 D_\nu {\pi_\mu}^\nu,
\]
where $D_\mu$ denotes the covariant derivative, and 
$\pi$ satisfies
\[
\{ g_{\mu\nu}(x), \pi^{\rho\sigma}(y) \}=\frac{1}{2} 
\left( \delta_\mu^\rho \delta_\nu^\sigma+ \delta_\mu^\sigma \delta_\nu^\rho\right)\delta^D(x-y).
\]
Then, for a vector $v$, we obtain
\[
\{J^{GR}(v),g(x)\}
&=\int d^Dy\ v^\mu (y) \{ J^{GR}_\mu(y),  g(x)\} \CR
&=\int d^Dy\ v^\mu (y) \{ -2 D_\nu {\pi_\mu}^\nu(y),g(x) \} \CR
&=2 \int d^Dy\ (D_\nu v^\mu (y)) \{  {\pi_\mu}^\nu(y),g(x) \} \CR
&=-2 g(x) D_\mu v^\mu(x). 
\label{eq:tJg}
\] 
Therefore,
\[
\{ J^{GR} (v), \tH(\xi)\}&=\int d^Dx\ \{ \cJ^{GR} (v), \xi(x) g(x)^{B} \cH^{GR}(x) \}  \CR
&=\int d^Dx 
\left( \xi(x) \{ J^{GR}(v),  g(x)^{B} \} \cH^{GR}(x) +\xi(x) g(x)^{B} \{ J^{GR}(v),\cH^{GR}(x) \} \right) \CR
&=\int d^Dx \left( -2 B \xi(x) g(x)^B D_\mu v^\mu(x) + v^\mu(x) \partial_\mu \left( \xi(x) g(x)^B\right) \right) 
\cH^{GR}(x),
\label{eq:tJtHmiddle}
\]
where we have used \eq{eq:tJg}, \eq{eq:grpoisson}, and \eq{eq:wn}.
Then, by substituting an identity,
\[
D_\mu v^\mu(x)&=\partial_\mu v^\mu(x)+\Gamma^\nu_{\nu\mu}(x) v^\mu(x) \CR
&=\partial_\mu v^\mu(x)+\frac{v^\mu(x)\partial_\mu g(x)}{2g(x)},
\]
into \eq{eq:tJtHmiddle}, we obtain
\[
\{ J^{GR}(v), \tH(\xi)\}
&=\int d^Dx  \ ( -2 B \xi(x)  \partial_\mu v^\mu(x) + v^\mu(x) \partial_\mu \xi(x) ) 
g(x)^B \cH^{GR}(x) \CR
&=\tH( -2 B \xi \partial_\mu v^\mu + v^\mu \partial_\mu \xi ).
\]
Thus, for $B=-\frac{1}{4}$, we certainly obtain \eq{eq:etaxi} of CTM.

\section{Summary and discussions}
\label{eq:summary}
The canonical tensor model (CTM) is a rank-three tensor model formulated as a totally constrained system 
with a number of first-class constraints, which have the Poisson algebraic structure similar to the constraint 
Poisson algebra of the ADM formalism of general relativity. 
In this paper, we consider a formal continuum limit of CTM and have shown that,
in the continuum limit, the constraint algebra of CTM coincides with that of the ADM formalism by
properly taking into account the weight difference:
the Hamiltonian constraints of CTM and the ADM formalism are different with each other by half-density.
We have obtained the expression of the (inverse) metric tensor 
field of general relativity in terms of the dynamical rank-three tensor $P$ of CTM.
Here the continuum limit assumes an almost diagonal form of $P$, and 
the lowest and the next to the lowest order coefficients of a moment expansion for the off-diagonal components of $P$
give the expression of the (inverse) metric tensor field. 
This explicit correspondence between $P$  and the metric tensor field would be the most useful 
achievement of this paper to guide future study of the dynamics of CTM.

A specific form of $P$, almost diagonal, is assumed in the continuum limit.
This is obviously an insufficient treatment, since $P$ is a dynamical variable, 
and its form must be determined dynamically rather than formally assumed. 
One possible way to justify it would be to study the exact physical wave functions which have been obtained 
so far \cite{Sasakura:2013wza,Narain:2014cya}, and check whether such configurations 
can appear as peaks of these wave functions. 
A non-trivial issue in such a study would be that we have to take a certain large $N$ limit, 
which would be necessary for indices to be replaced by continuous coordinates.

We have seen that the cosmological constant of CTM generates non-local dynamics which 
is in contradiction with the locality assumption of the continuum limit. 
Therefore, for the consistency of our discussions, the cosmological constant 
must vanish. 
In fact, we have previously shown that the value of the cosmological constant can 
be changed by shifting $P$ \cite{Narain:2014cya}. 
It would be interesting to study the exact physical wave functions 
\cite{Sasakura:2013wza,Narain:2014cya} to see whether there is a dynamical mechanism which 
tunes $P$ to cancel the cosmological constant. 
Note that the dynamics of the cosmological constant term of CTM is obviously different from 
that of general relativity due to the non-local property, and this would give a certain 
chance to circumvent the common difficulties to tune the cosmological constant to the 
observed value \cite{Weinberg:1988cp}.
Another possibility is that the cosmological constant term is prohibited by the consistency of the 
constraint algebra of CTM. As shown in a previous paper \cite{Sasakura:2012fb}, this actually occurs, 
if we impose the generalized Hermiticity condition on the dynamical variables of CTM, 
instead of imposing the reality condition as in this paper.     
 
We have chosen a local class of momentum constraints of CTM from all, 
and have shown that they correspond to the momentum constraints in the ADM formalism. 
However, we have not discussed consequences of ignoring the non-local momentum constraints of CTM. 
One way to justify the ignorance would be 
to gauge-fix the non-local ones, while the local ones are left intact.
Then, a consequence in quantum case would be that a constraint, say $\hat {\cal C}$, 
would be modified by a similarity transformation $\hat {\cal C}_{\text{eff}}=\sqrt{V} \hat {\cal C} \sqrt{V}^{-1}$
\cite{Jevicki:1979mb}, where $V$ is the gauge volume of the non-local gauge transformations.
In a more general treatment, $V$ could also contain a Jacobian generated from 
a process of taking the coefficients of the moment expansions
as coarse grained collective coordinates for $P$ \cite{Jevicki:1979mb}.
Such a similarity transformation does not change the algebraic structure, but
there will be some consequences of physical importance. 
One is that an almost diagonal form will be required for 
$\hat P_{\text{eff}}=\sqrt{V} \hat P \sqrt{V}^{-1}$ instead of $\hat P$ in the continuum limit. 
Another will be that Hamiltonian constraint 
$\hat {\cal H}_{\text{eff}}=\sqrt{V} \hat {\cal H} \sqrt{V}^{-1}$ will be changed from the original
form $\hat {\cal H}$. Since the original Hamiltonian constraint of CTM seems lacking a term
corresponding to the spatial curvature term in that of the ADM formalism, 
it would be highly interesting to see whether
the corrections will modify the Hamiltonian constraint in a desired manner or not.

\section*{Acknowledgements} 
NS was supported in part by JSPS KAKENHI Grant Number 15K05050.
NS would especially like to thank C.~Rovelli and his group members for 
their hospitality and stimulating discussions during his stay in CPT, Marseille.  
NS would also like to thank V.~Rivasseau for hospitality and some discussions 
during his stay in LPT, Paris, and 
M.~Fukuma for stimulating discussions on recent developments of related subjects.  
YS would like to thank J.~Rodrigues and A.~Tsuchiya for useful communications.

\end{document}